\documentstyle[twocolumn,seceq,epsbox]{jpsj}

\def\runtitle{Field-Induced Two-Step Phase Transitions in CsFeBr$_3$}
\def\runauthor{Yosuke {\sc Tanaka} {\it et al.}}

\title
{
Field-Induced Two-Step Phase Transitions in the Singlet Ground State Triangular Antiferromagnet CsFeBr$_3$
}

\author
{ 
Yosuke {\sc Tanaka}, Hidekazu {\sc Tanaka}, Toshio {\sc Ono}, Akira {\sc Oosawa}, \\
Kiyoko {\sc Morishita}, Katsunori {\sc Iio}, Tetsuya {\sc Kato}$^{1}$, Hiroko {\sc Aruga Katori}$^{2}$, \\
Mikhail I. {\sc Bartashevich}$^{3}$ and Tsuneaki {\sc Goto}$^{3}$
}

\inst
{
Department of Physics, Tokyo Institute of Technology, Oh-okayama, Meguro-ku, Tokyo 152-8551\\
$^1$Faculty of Education, Chiba University, Inage-ku, Chiba 263-8522\\
$^2$RIKEN (The Institute of Physical and Chemical Research), Wako, Saitama 351-0198\\
$^3$Institute for Solid State Physics, The University of Tokyo, Kashiwanoha, Kashiwa, Chiba 277-8581
}

\recdate
{
\hspace{6cm}
}

\abst
{
The ground state of the stacked triangular antiferromagnet CsFeBr$_3$ is a spin singlet due to the large single ion anisotropy $D(S^z)^2$. The field-induced magnetic ordering in this compound was investigated by the magnetic susceptibility, the magnetization process and specific heat measurements for an external field parallel to the $c$-axis. Unexpectedly, two phase transitions were observed in the magnetic field $H$ higher than 3 T. The phase diagram for temperature versus magnetic field was obtained. The mechanism leading to the successive phase transitions is discussed.}

\kword
{
CsFeBr$_3$, triangular antiferromagnet, singlet ground state, field-induced magnetic ordering, successive phase transitions, single ion anisotropy, magnetic susceptibility, magnetization process, specific heat
}

\begin{document}
\sloppy
\maketitle

\section{Introduction}
The spin frustration effect often plays an important role in the magnetic ordering process and magnetic excitations. In the hexagonal antiferromagnets of ABX$_3$-type with the CsNiCl$_3$ structure, magnetic B$^{2+}$ ions form infinite chains along the $c$-axis and triangular lattices in the basal $c$-plane. Since the exchange interaction in the $c$-plane is antiferromagnetic, they behave as triangular antiferromagnets (TAF) at low temperatures. Because of the spin frustration effect being characteristic of TAF, together with the quantum effect, a rich variety of phase transitions have been observed in the hexagonal ABX$_3$ antiferromagnets.\cite{Collins} 

Low-temperature magnetic properties of AFeX$_3$ systems are described by the pseudo spin $S=1$ anisotropic $XXZ$ model with the large easy-plane anisotropy of the form $D(S^z)^2$ due to the crystalline field.\cite{Yoshizawa} In RbFeCl$_3$, exchange interactions overcome the anisotropy, so that RbFeCl$_3$ undergoes magnetic phase transition in the absence of magnetic field. On the other hand, CsFeCl$_3$ has a singlet ground state at zero field, because exchange interactions are not sufficiently strong to produce the magnetic ordering. The ordering process in RbFeCl$_3$ is not simple. The phase transition occurs from the paramagnetic state to the commensurate (C) ground state with the 120$^{\circ}$ structure through two different incommensurate (IC) states.\cite{Haseda,Wada1} The IC-C phase transition occurs due to the competition between the antiferromagnetic exchange interaction in the $c$-plane and the dipole-dipole (D-D) interaction, the latter of which is enhanced by the ferromagnetic exchange interaction along the $c$-axis.\cite{Shiba1,Shiba2} A similar IC-C phase transition was observed in CsFeCl$_3$ when magnetic fields $H$ are applied parallel to the $c$-axis.\cite{Haseda,Knop,Chiba} In addition, in RbFeCl$_3$, the quantum fluctuation produces successive phase transitions in magnetic fields perpendicular to the $c$-axis.\cite{Wada2,Ono,Shiba3} 
\begin{table}[t]
\caption{Interaction parameters for CsFeBr$_3$ in the unit of K.}
\label{table:1}
\begin{tabular}{@{\hspace{\tabcolsep}\extracolsep{\fill}}cccccc} \hline
$D/k_{\rm B}$ & $J_0^{\parallel}/k_{\rm B}$ & $J_0^{\perp}/k_{\rm B}$ & $J_1^{\parallel}/k_{\rm B}$ & $J_1^{\perp}/k_{\rm B}$ & ref.   \\ \hline
29.8 & 3.2 & 3.2 & 0.32 & 0.32 & \cite{Dorner1} \\
21.4 & 4.9 & 4.4 & 0.48 & 0.43 & \cite{Visser1,Harrison} \\
\hline
\end{tabular}
\end{table}
CsFeBr$_3$ is isostructural with RbFeCl$_3$ and CsFeCl$_3$.\cite{Takeda} However, CsFeBr$_3$ differs from them in terms of the exchange interaction along the $c$-axis. CsFeBr$_3$ has an antiferromagnetic intrachain interaction. Thus, the D-D interaction should be negligible. As observed in CsFeCl$_3$, the ground state of CsFeBr$_3$ is a spin singlet due to the large single ion anisotropy.\cite{Bocquet,Dorner1} The effective Hamiltonian of CsFeBr$_3$ at low temperatures for $H\parallel c$ may be written as
\begin{eqnarray}
{\cal H} &=& \sum_{i}\left[\,D(S_{i}^z)^2-g_{\parallel}\mu_{\rm B}HS_{i}^z\,\right] \nonumber \\
         & & {} +\sum_{\langle i,j\rangle}^{\rm chain}\left[J_0^{\perp}(S_i^+S_j^- + S_i^-S_j^+) + 2J_0^{\parallel}S_i^zS_j^z\right] \nonumber\\
         & & {} +\sum_{\langle i,j\rangle}^{\rm plane}\left[J_1^{\perp}(S_i^+S_j^- + S_i^-S_j^+) + 2J_1^{\parallel}S_i^zS_j^z\right]  , \nonumber \\
         & &
\end{eqnarray}
where $S_i^{\alpha}\ (\alpha=x,y,z)$ is the spin-1 operator on the $i$-th Fe$^{2+}$ site, and $J_0^{\parallel, \perp}$ and $J_1^{\parallel, \perp}$ are the antiferromagnetic exchange interactions along the $c$-axis and in the $c$-plane, respectively.

Magnetic excitations in CsFeBr$_3$ have been extensively investigated by means of neutron inelastic scattering experiments at zero and finite magnetic fields.\cite{Dorner1,Visser1,Dorner2,Visser2,Schmid,Harrison} The lowest excitation occurs at $\mib Q=(1/3, 1/3, 1)$ and its equivalent reciprocal points with the excitation energy of $\Delta=0.11$ THz.\cite{Schmid} The results were analyzed on the basis of the Hamiltonian of eq. (1.1), and the interaction parameters were determined, as listed in Table I. 

When an external field $H$ is applied along the $c$-axis, CsFeBr$_3$ undergoes a magnetic phase transition for $H>4$ T.\cite{Dorner2} The magnetic structure in the ordered phase is a triangular structure characterized by an ordering vector $\mib Q=(1/3, 1/3, 1)$, which was at first assumed to be the 120$^\circ$ spin structure in the basal plane.\cite{Dorner2,Lindgard} 
\begin{figure}[t]
%\figureheight{5cm}
    \epsfigure{file=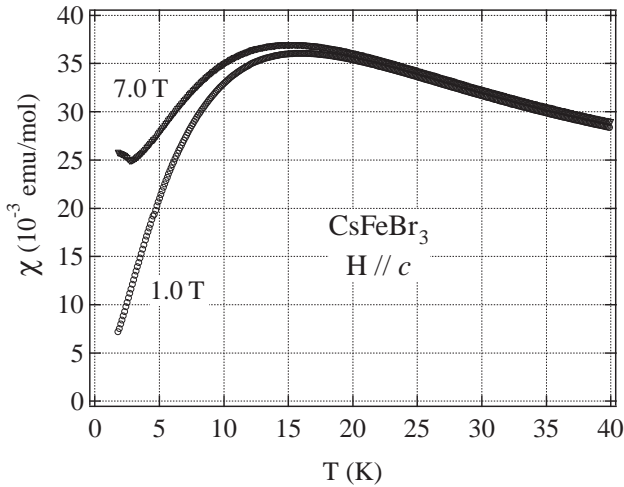,width=90mm}
\caption{The temperature dependence of the magnetic susceptibility $M/H$ in CsFeBr$_{3}$ at 1 and 7 T for $H\parallel c$.}
\label{fig:1}
\end{figure}
The stacked TAF with $XY$ spins exhibits a phase transition from the paramagnetic state to the ordered state with the 120$^\circ$ spin structure. Recently, the nature of this phase transition has been under controversy. Kawamura\cite{Kawamura1,Kawamura2} argued that the phase transition is of the second order, and belongs to a new chiral $XY$ universality class. His theory was supported by neutron scattering and specific heat experiments performed on CsMnBr$_3$.\cite{Kadowaki,Ajiro,Mason,Deutschmann} However, it was shown later that the phase transition must be of the first order.\cite{Plumer,Loison} This problem motivated us to study the field-induced magnetic ordering in CsFeBr$_3$. Since CsFeBr$_3$ has a large easy-plane anisotropy, which overcomes all exchange interactions, the field-induced phase transition for $H\parallel c$ may have the same nature as that of the $XY$ stacked TAF. In order to investigate the field-induced magnetic phase transition in CsFeBr$_3$, we have performed magnetization and specific heat measurements. As shown below, unexpected two-step phase transitions were observed. 

The arrangement of this paper is as follows: In \S 2, the experimental procedures are described. The experimental results and their analyses are presented in \S 3. Sections 4 and 5 are devoted to the discussion and conclusion, respectively. 

\begin{figure}[t]
%\figureheight{12cm}
    \epsfigure{file=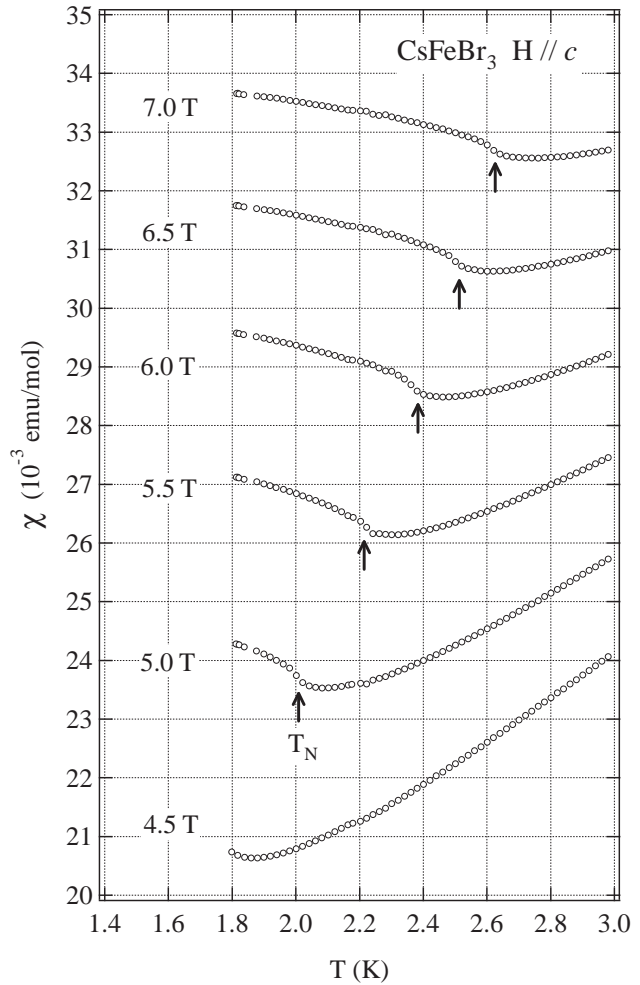,width=90mm}
\caption{The low-temperature magnetic susceptibility in CsFeBr$_{3}$ at various external fields for $H\parallel c$.}
\label{fig:2}
\end{figure}

\section{Experimental Procedures}
Single crystals of CsFeBr$_3$ were grown by the vertical Bridgman method from the melt of an equimolar mixture of CsBr and FeBr$_2$ sealed in evacuated quartz tubes. After weighing, they were placed in a quartz tube and dehydrated by heating in vacuum at $\simeq$150$^{\circ}$C for three days. The temperature at the center of the furnace was set at 650$^{\circ}$C, and the lowering rate was 3 mm$\cdot$h$^{-1}$. Single crystals of 1$\sim$5 cm$^3$ were obtained. The crystals obtained were identified to be CsFeBr$_3$ by X-ray powder diffraction. 

The specific heat measurements for single crystal of CsFeBr$_3$ were carried out at RIKEN down to 0.6 K in magnetic fields up to 10 T using a Mag Lab$^{\rm HC}$ microcalorimeter (Oxford Instruments) in which the relaxation method was employed.
The magnetizations were measured down to 1.8 K in magnetic fields up to 7 T using a SQUID magnetometer (Quantum Design MPMS XL). 
The high-field magnetization measurement was performed using an induction method with a multilayer pulse magnet at the Ultra-High Magnetic Field Laboratory, Institute for Solid State Physics, The University of Tokyo. Magnetization data were collected at 1.6 K in magnetic fields up to 40 T. 
In these experiments, magnetic fields were applied along the $c$-axis. 
\begin{figure}[t]
%\figureheight{5cm}
    \epsfigure{file=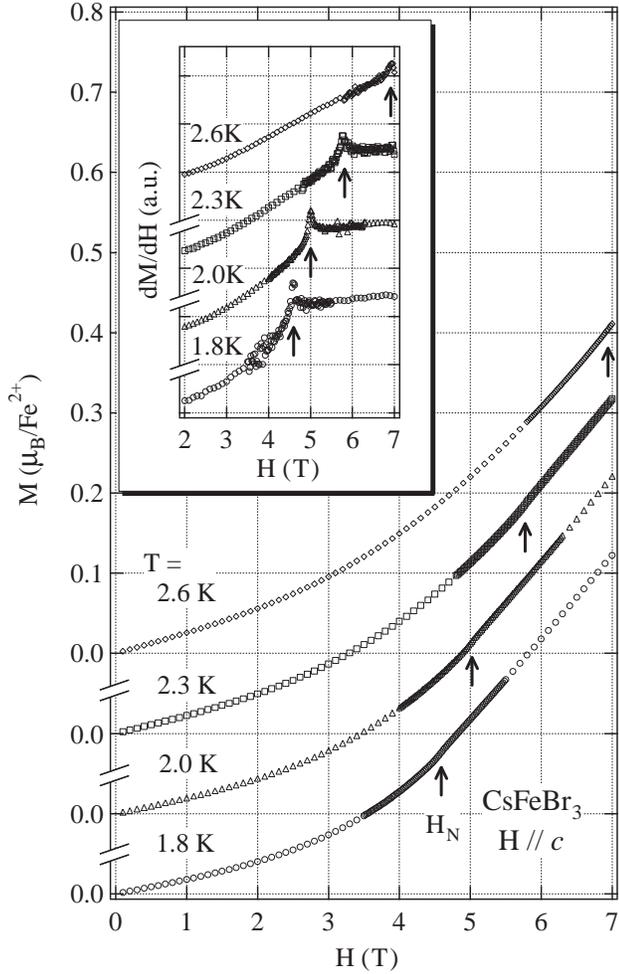,width=90mm}
\caption{Magnetization curves in CsFeBr$_{3}$ at various temperatures for $H\parallel c$. The inset shows $dM/dH$ versus $H$.}
\label{fig:3}
\end{figure}
\section{Experimental Results}
\subsection{Magnetic susceptibility and low-field magnetization}
We first measured the magnetic susceptibilities in CsFeBr$_3$ at $H=1$ and 7 T for $H\parallel c$. The results are shown in Fig. 1. With decreasing temperature, the susceptibility $\chi$ ($=M/H$) exhibits a broad maxima at about 15 K and then decreases. The susceptibility for $H=1$ T decreases monotonically toward zero, while the susceptibility for $H=7$ T displays a cusplike minimum at $T=2.62$ K, which indicates the magnetic phase transition. 
\begin{figure}[t]
%\figureheight{12cm}
    \epsfigure{file=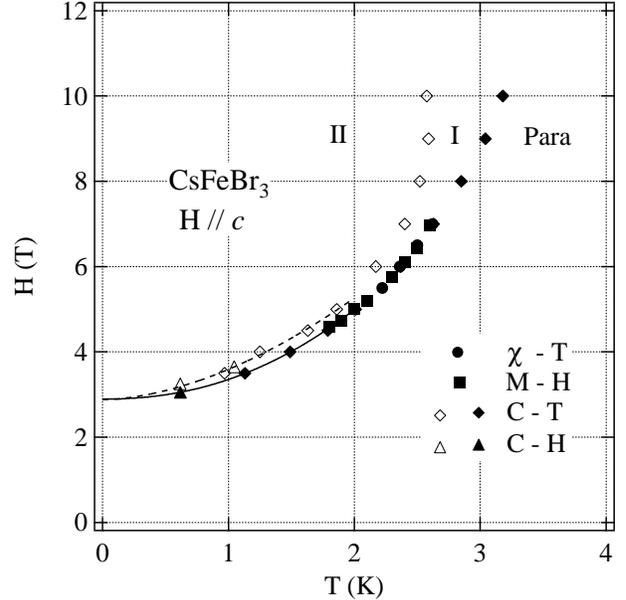,width=90mm}
\caption{The phase boundary in CsFeBr$_{3}$ for $H\parallel c$ determined from the results of the temperature variations and field variations of the magnetization and the specific heat. Solid circles and rectangles denote the transition points determined from the temperature dependence of susceptibilities and the magnetization curves, respectively. Diamonds and triangles mean the transition points determined from the temperature and field variations of specific heat, respectively. Solid and dashed lines are the fits with eq. (3.1) with $\phi_1=2.2(1)$ and $H_{\rm N1}(0)=2.89(2)$ T, and $\phi_2=1.8(1)$ and $H_{\rm N2}(0)=2.88(2)$ T, respectively.}
\label{fig:4}
\end{figure}
Figure 2 shows the low-temperature susceptibility in CsFeBr$_3$ measured at various external fields for $H\parallel c$. We assign the temperature at which there is an inflection point in the susceptibility as the transition temperature $T_{\rm N}(H)$. The arrows in Fig. 2 indicate the transition temperatures, which are slightly lower than the temperatures of the susceptibility minima. With increasing external field, the transition temperature increases. 

Figure 3 shows the magnetization curves in CsFeBr$_3$ measured at various temperatures for $H\parallel c$. $dM/dH$ versus $H$ is presented in the inset. A sharp peak anomaly, which is indicative of the phase transition, can be observed in $dM/dH$ versus $H$. We assign the field at which $dM/dH$ exhibits the peak anomaly as the transition field $H_{\rm N}(T)$. The arrows in Fig. 3 indicate the transition fields. With increasing temperature, the transition field increases. In spite of the singlet ground state, the slope of the magnetization curve for $H<H_{\rm N}$ is fairly large even at $T=1.8$ K. This is because of the large Van Vleck paramagnetism due to the crystalline field.

In the present magnetization measurements, a single phase transition was clearly observed. In Fig. 4, the phase transition temperatures $T_{\rm N}(H)$ and fields $H_{\rm N}(T)$ were plotted as solid circles and rectangles, respectively. Because the transition points determined from $T_{\rm N}(H)$ and $H_{\rm N}(T)$ lie almost on the same line, they are consistent with each other.

\begin{figure}[t]
%\figureheight{12cm}
    \epsfigure{file=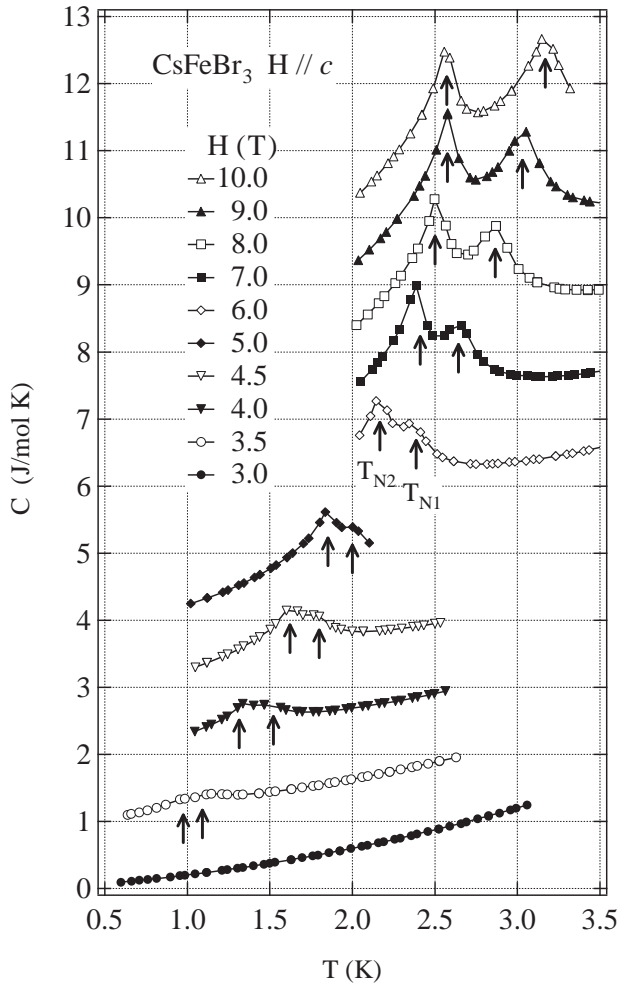,width=90mm}
\caption{The temperature dependence of the specific heat in CsFeBr$_{3}$ at various magnetic fields for $H\parallel c$. The values of the specific heat are shifted upward by 1 J/mol K with increasing external field.}
\label{fig:5}
\end{figure}

\subsection{Specific heat}
Figure 5 shows the total specific heat $C$ in CsFeBr$_3$ measured at various external fields for $H\parallel c$. For $H=3$ T, no anomaly is observed down to $T=0.6$ K, while for $H=3.5$ T, the anomaly composed of two peaks can be observed. With increasing external field, the two peaks shift toward the high-temperature side, and separate increasingly. They are well defined for $H\geq 7$ T. We confirmed that the sample had not been cracked into two pieces during the cooling, and that the values of the specific heat are reproducible using another sample. Thus, the two-peak anomaly in $C$ is not due to a sample problem, but is intrinsic to the present system. Based on the result of the specific heat measurements, we can conclude that the magnetic ordering in CsFeBr$_3$ occurs in two steps. The arrows in Fig. 5 denote the phase transition temperatures $T_{\rm N1}$ and $T_{\rm N2}$.

In order to determine the phase transition points for $H<3.5$ T, which are difficult to determine from the temperature scan, we measured the field dependence of $C$ for $H\parallel c$.  Figure 6 shows the measurements at $T=0.62$ K. A shoulder and cusplike anomalies are visible at $H_{\rm N1}=3.05$ T and $H_{\rm N2}=3.25$ T. 
\begin{figure}[t]
%\figureheight{12cm}
    \epsfigure{file=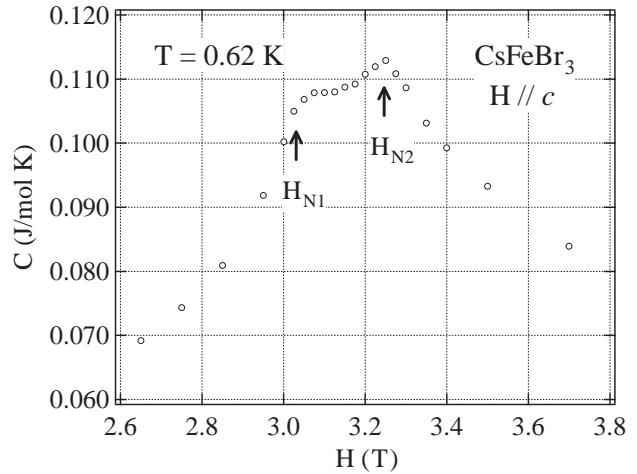,width=90mm}
\caption{The field dependence of the specific heat in CsFeBr$_{3}$ at 0.62 K for $H\parallel c$.}
\label{fig:6}
\end{figure}
In Fig. 4, the phase transition points determined from the specific heat measurements are plotted as diamonds and triangles. It is evident that the outer phase boundary determined from $T_{\rm N1}$ and $H_{\rm N1}$ coincides with the phase boundary determined from the magnetization measurements. The temperature interval of the intermediate phase increases with the magnetic field. We label the intermediate phase and the low-temperature phase inside the inner boundary phases I and II, respectively. 
\begin{fullfigure}
\begin{minipage}{8.5cm}
\begin{center}
\epsfigure{file=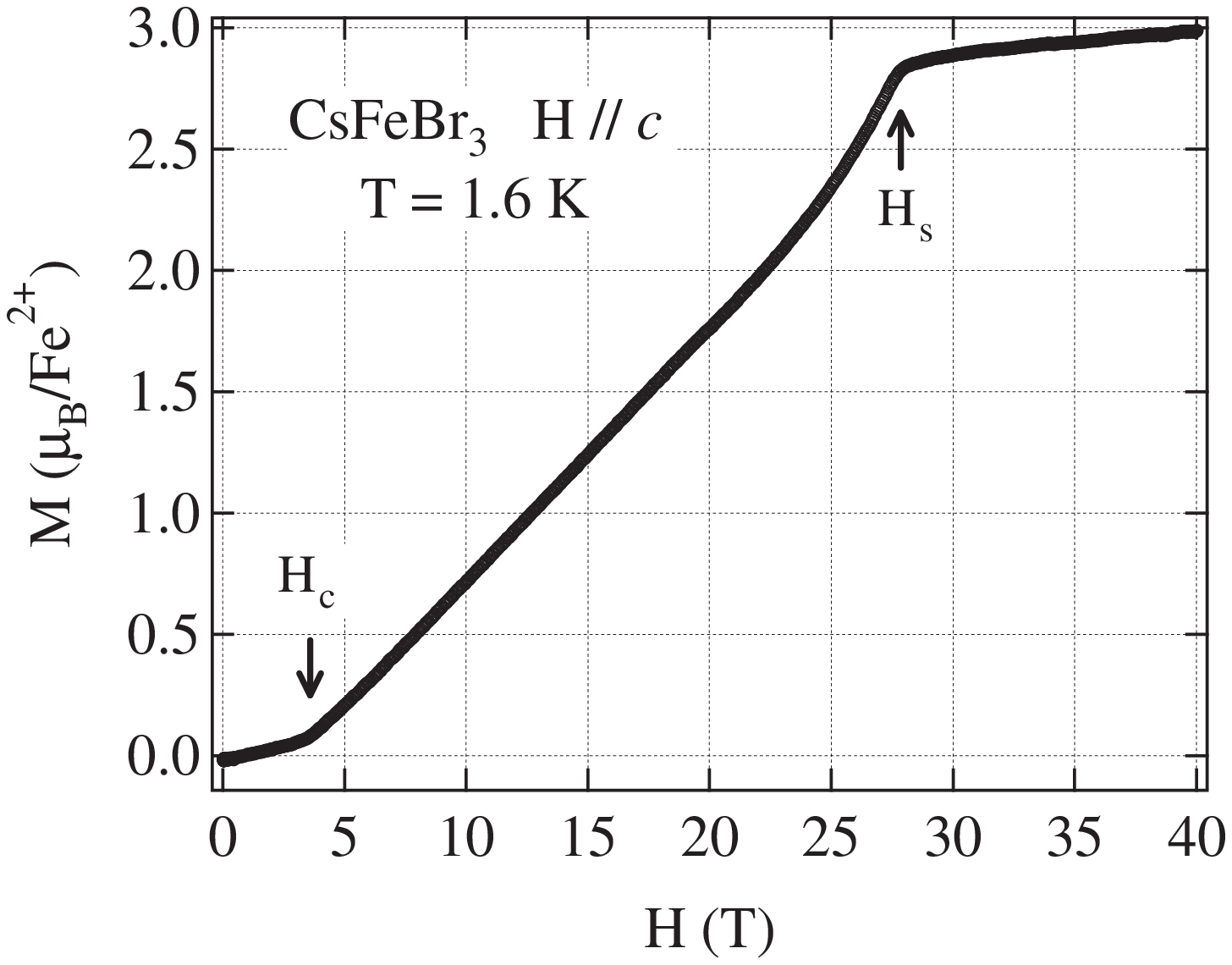,width=90mm}
(a)
\end{center}
\end{minipage}
\begin{minipage}{8.5cm}
\begin{center}
\epsfigure{file=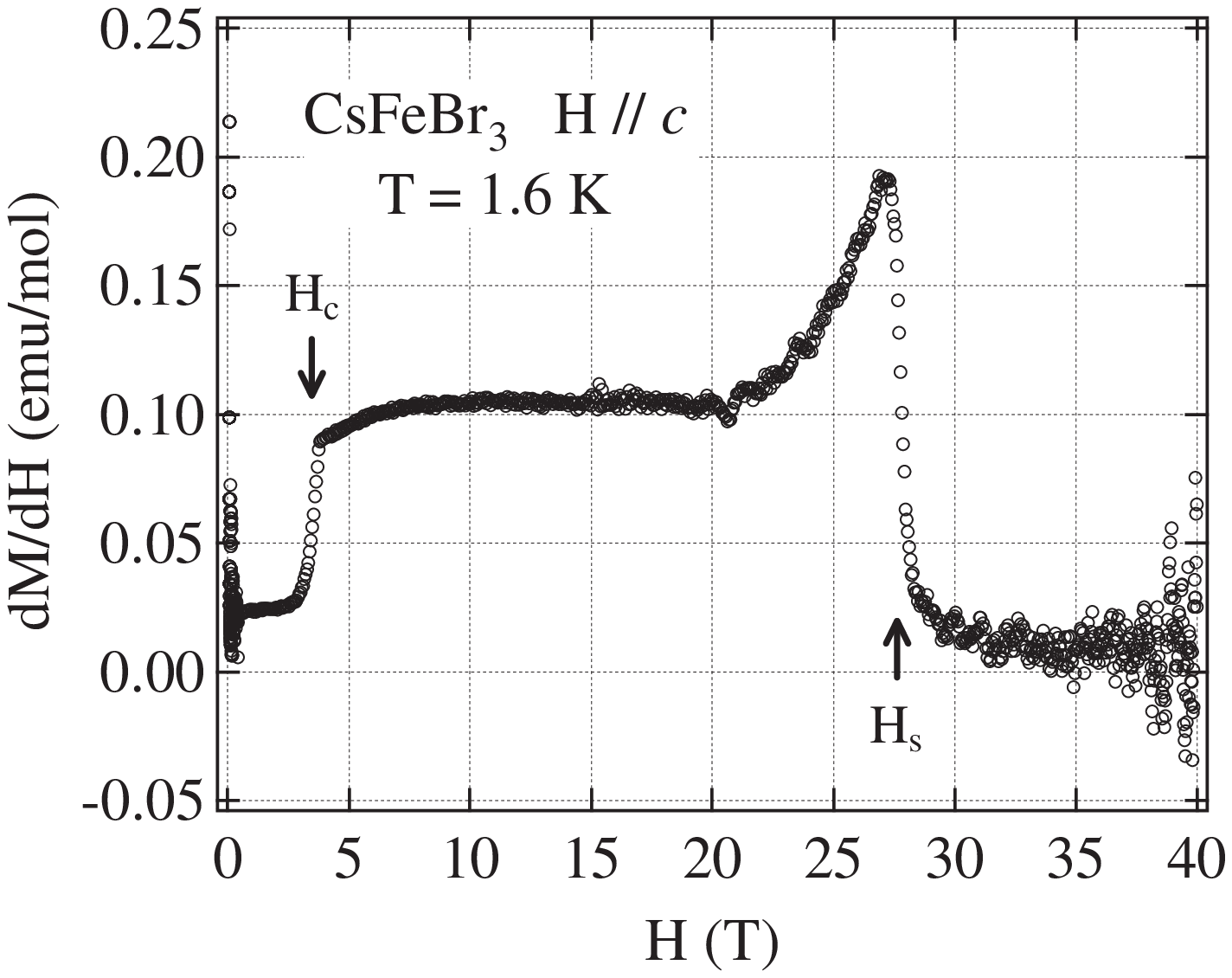,width=90mm}
(b)
\end{center}
\end{minipage} \\
\caption{(a) Magnetization curve and (b) $dM/dH$ vs $H$ in CsFeBr$_{3}$ at 1.6 K for $H\parallel c$.}
\label{fig:7}
\end{fullfigure}
When the magnetic field is slightly higher than the gap field $H_{\rm g}$ corresponding to the excitation gap, both phase boundaries may be represented by the power law
\begin{eqnarray}
\left[H_{{\rm N}\alpha}(T)-H_{{\rm N}\alpha}(0)\right] \propto T^{\phi_\alpha} ,
\end{eqnarray}
with $\alpha=$1 and 2. We fit eq. (3.1) to the data for $T<2$ K, which is lower than half the gap temperature $\Delta/k_{\rm B}\approx 5.3$ K.\cite{Schmid} Solid and dashed lines in Fig. 4 are fits with $\phi_1=2.2(1)$ and $H_{\rm N1}(0)=2.89(2)$ T, and $\phi_2=1.8(1)$ and $H_{\rm N2}(0)=2.88(2)$ T, respectively. Since $H_{\rm N1}(0)\approx H_{\rm N2}(0)$, we can conclude that the two phase boundaries meet at zero temperature. Thus, the gap field $H_{\rm g}$ for $T=0$ is estimated to be $H_{\rm g}=H_{\rm N1}(0)=2.89$ T.

\subsection{High-field magnetization}
In order to determine the phase boundaries in high fields, we performed high-field magnetization measurements. Figure 7 shows the magnetization curve and $dM/dH$ versus $H$ measured at $T=1.6$ K. Sharp anomalies indicative of phase transitions are observed at $H_{\rm c}=3.54$ T and $H_{\rm s}=27.8$ T. The transition field $H_{\rm c}$ should correspond to $H_{\rm N1}$ observed through low-field magnetization and specific heat measurements. $H_{\rm s}$ is the saturation field. 

The magnetization curve for $H>H_{\rm s}$ has a finite slope due to the large Van Vleck paramagnetism, {\it i.e.}, $\chi_{\rm VV}=4.64{\times}10^{-3}$ emu/mol. The saturation magnetization is evaluated to be $M_{\rm s}=2.65\pm 0.10$ $\mu_{\rm B}$/Fe$^{2+}$ by extrapolating the magnetization curve for $H>H_{\rm s}$ to zero field. The value of $M_{\rm s}$ is almost the same as $M_{\rm s}=2.6$ $\mu_{\rm B}$/Fe$^{2+}$ in CsFeCl$_3$.\cite{Chiba} From the value of $M_{\rm s}$, the $g$-factor for $H\parallel c$ is obtained to be $g_{\parallel}=2.65\pm 0.10$. The $g$-factor can also be evaluated to be $g_{\parallel}=2.7$ using the relation of $\Delta=g\mu_{\rm B}H_{\rm g}$ with $\Delta=0.11$ THz\cite{Schmid} and $H_{\rm g}=2.89$ T. Both $g$-factors coincide and agree with previously obtained $g_{\parallel}=2.6$.\cite{Visser2}

The transition points obtained from the high-field magnetization measurements are plotted as solid triangles in Fig. 8 together with those obtained from the low-field magnetization and specific heat measurements. The solid lines are guides for the eyes. The dashed lines denote the expected phase boundaries.

\section{Discussion}
The field-induced magnetic ordering in the singlet ground state magnet with the large $D$ term was first investigated by Tsuneto and Murao\cite{Tsuneto} on the basis of the mean-field theory. They demonstrated that a single-ordered phase exists in a closed area in the magnetic field vs temperature ($H-T$) diagram. Such field-induced phase transition was observed in several nickel compounds, {\it e.g.}, ${\rm Ni(C_5H_5NO)_6(ClO_4)_2}$,\cite{Diederix,Algra} ${\rm Ni(NO_3)_2\cdot 4H_2O}$\cite{Wada3} and ${\rm Ni(NO_3)_2\cdot 6H_2O}$,\cite{Wada4} all of which are unfrustrated systems. 

Since the intrachain exchange interaction $J_0$ is antiferromagnetic in CsFeBr$_3$, the dipole-dipole interaction should be negligible. Thus, it has been considered that CsFeBr$_3$ would undergo a single phase transition from the paramagnetic phase to the ordered phase with the 120$^\circ$ spin structure in the basal plane in high magnetic fields. However, the present study revealed that the field-induced phase transition occurs in two steps. This is an unexpected finding. The phase diagram for $H\parallel c$ was determined as shown in Figs. 4 and 8. 

The magnetic ordering in CsFeBr$_3$ was investigated by means of neutron elastic scattering in magnetic fields up to 6 T.\cite{Dorner2,Visser3} The magnetic Bragg reflections were observed at $\mib Q=(1/3, 1/3,1)$ and $(2/3, 2/3,1)$. Thus, it is certain that a triangular spin structure is realized in the basal plane in the phase II. However, it was also suggested from the magnetic excitations in phase II that the triangular spin structure is not the regular 120$^{\circ}$ structure but is a slightly modified one.\cite{Visser3} The transition fields determined from the field dependence of the Bragg reflections for $T\leq 1.6$ K appear to be on the outer phase boundary in Fig. 4 (or Fig. 8). The distinct anomaly that is indicative of the second transition was not observed in the field variation of the magnetic Bragg reflections. Since the field interval of the intermediate phase I is less than 0.3 T for $T\leq 1.6$ K, it seems difficult to distinguish phase I.
\begin{figure}[t]
%\figureheight{12cm}
    \epsfigure{file=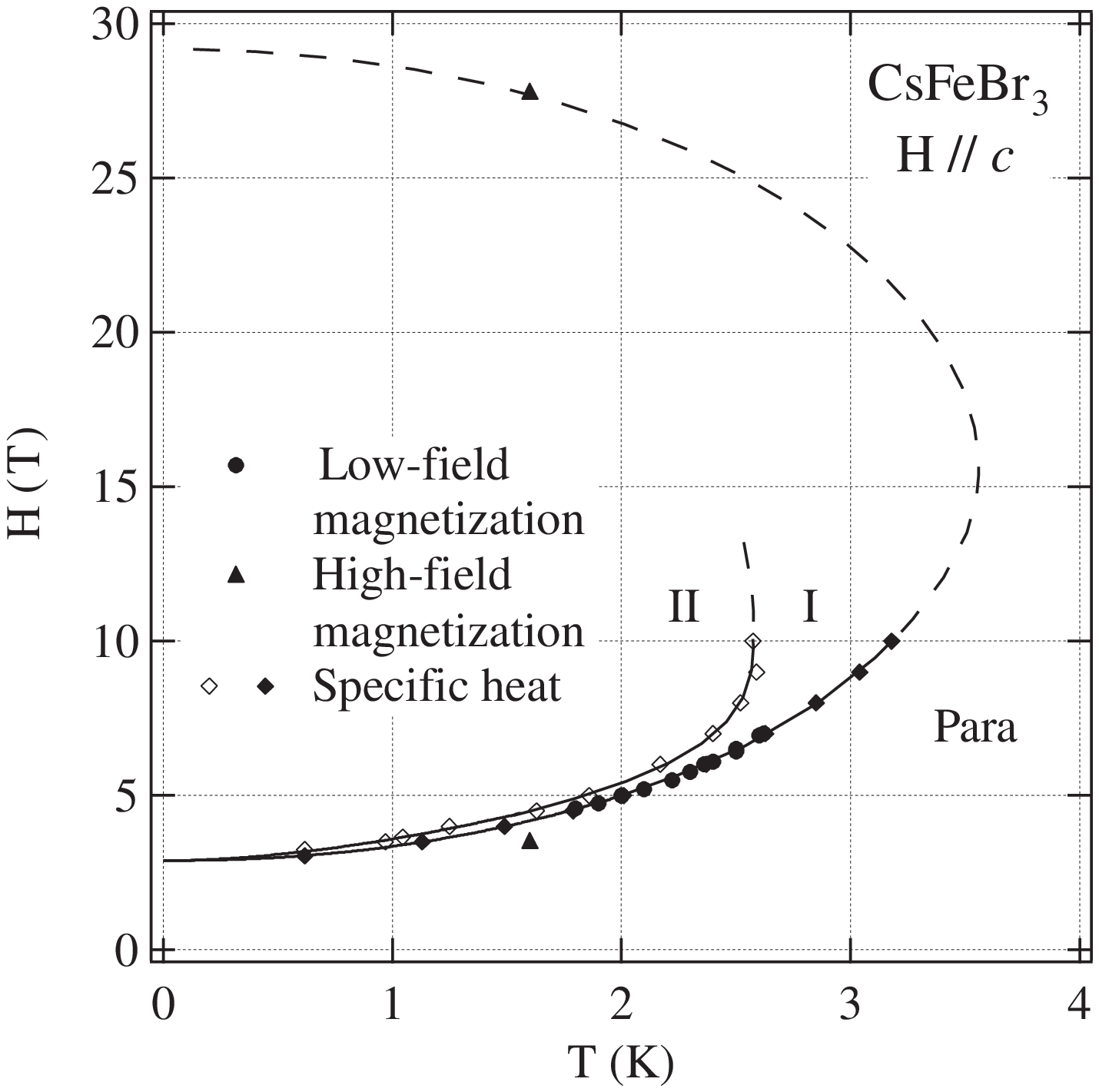,width=90mm}
\caption{The phase diagram in CsFeBr$_{3}$ for $H\parallel c$ determined by the present measurements. The solid lines are guides for the eyes. The dashed lines denote the expected phase boundaries.}
\label{fig:8}
\end{figure}
In the closely related compound RbFeBr$_3$, the total of the exchange interactions overcomes the single ion anisotropy energy (see eq. (A.13) in the Appendix), so that the magnetic ordering occurs even at zero field. RbFeBr$_3$ undergoes two magnetic phase transitions at $T_{\rm N1}=5.61$ K and $T_{\rm N2}=2.00$ K at zero field.\cite{Adachi} This compound has two structural phase transitions at $T_{\rm I}=109$ K and $T_{\rm C}=34.4$ K, which are accompanied with the shift of -FeBr$_3$- chains along the $c$-axis.\cite{Eibschutz,Mitsui} The chemical unit cell below $T_{\rm I}$ is described by enlarging that of CsNiCl$_3$ to $\sqrt3a$, $\sqrt3a$, $c$. Consequently, there exist two different interchain exchange interactions, $J_1$ and $J'_1$, which are equivalent in the undistorted CsNiCl$_3$ structure. The successive magnetic phase transitions in RbFeBr$_3$ arise from $J_1\neq J'_1$,\cite{Adachi} which are explainable in the framework of the mean-field theory. A collinear and a triangular spin structures are realized in the intermediate- and the low-temperature phases, respectively. The successive magnetic phase transitions due to the same mechanism were also observed in RbVBr$_3$.\cite{Tanaka2} 
\begin{figure}[t]
%\figureheight{12cm}
    \epsfigure{file=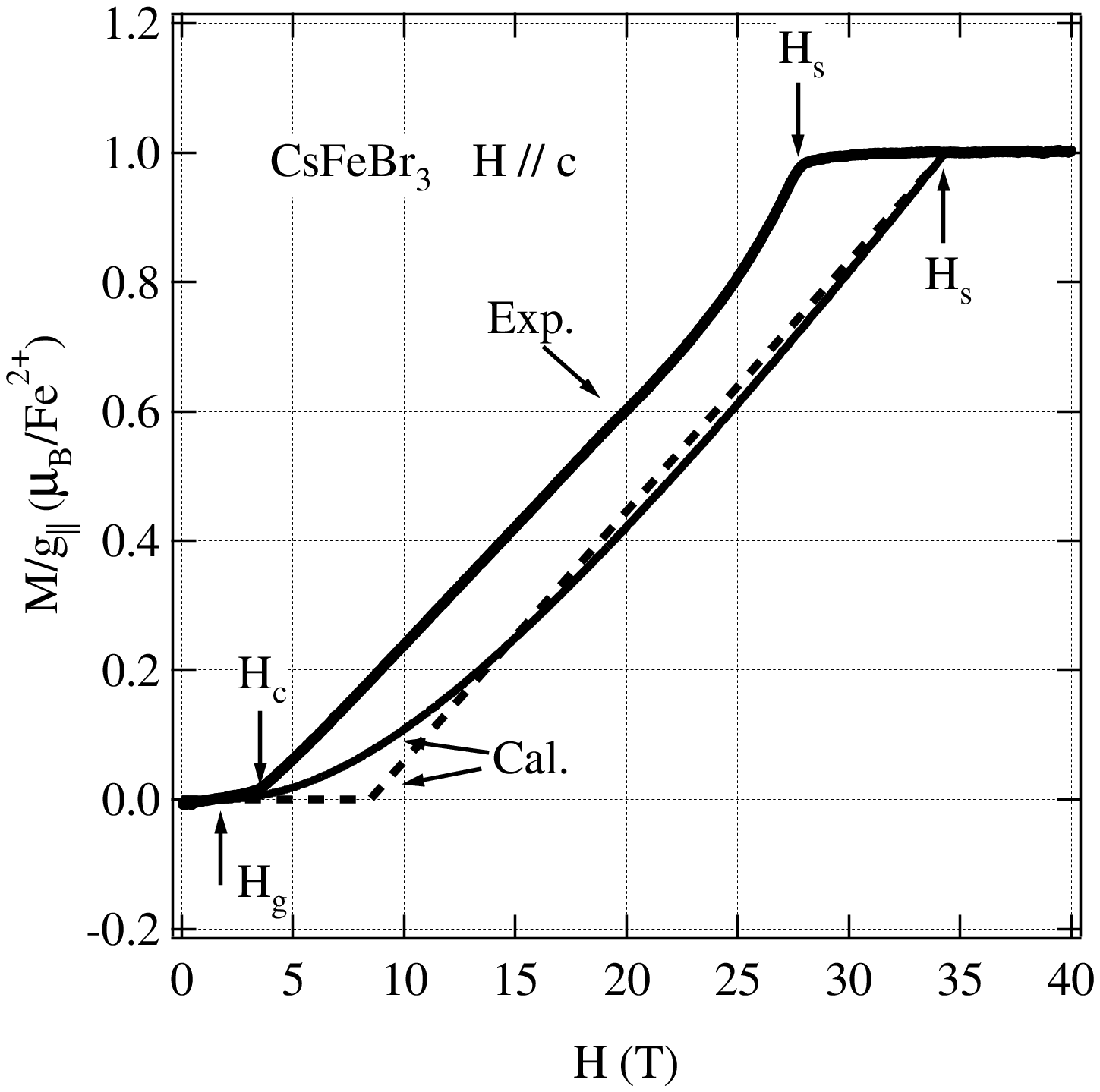,width=90mm}
\caption{Magnetization curve for $H\parallel c$ calculated by the mean-field approximation with $g_{\parallel}=2.65$ and the magnetic parameters obtained by Dorner {\it et al}.\cite{Dorner1} (see Table I). The solid and dashed lines are the results in the cases when the highest $|-1\,\rangle$ state is included and excluded, respectively. The thick solid line is the magnetization curve observed in CsFeBr$_{3}$ at $T=1.6$ K.}
\label{fig:9}
\end{figure}
Such structural phase transition has often been observed in hexagonal RbBBr$_3$ compounds, {\it e.g.}, RbFeBr$_3$,\cite{Eibschutz,Mitsui} RbVBr$_3$,\cite{Hauser} RbMnBr$_3$\cite{Kato} and RbCoBr$_3$,\cite{Morishita} but not in CsBBr$_3$ compounds except for the Jahn-Teller system CsCrBr$_3$.\cite{Li} CsFeBr$_3$ has the same crystal structure as CsNiCl$_3$ (space-group $P6_3/mmc$) at room temperature,\cite{Takeda} which we confirmed by X-ray powder diffraction. In order to check whether there is a structural phase transition in CsFeBr$_3$, we measured the dielectric constant between 4.2 K and 300 K at zero magnetic field. However, no anomaly indicative of a structural phase transition was detected. Thus, the crystal structure remain highly symmetric down to helium temperatures. Therefore, the successive field-induced magnetic phase transitions in CsFeBr$_3$ should not be attributed to $J_1\neq J'_1$ due to the structural phase transition. However, there is a possibility that the magnetoelastic coupling produces the situation $J_1\neq J'_1$ below $T_{\rm N1}$ in magnetic fields.

The two-step transitions observed in CsFeBr$_3$ cannot be described within the framework of the mean-field theory for the undistorted crystal structure, because it predicts a single phase transition. In the triangular antiferromagnet with the singlet ground state as CsFeBr$_3$, both the spin frustration and the quantum fluctuation may play important roles in the field-induced magnetic phase transition. Thus, not only the transverse spin fluctuation, but also the longitudinal spin fluctuation, which can be neglected in the classical spin system, will be responsible for the ordering process. We infer that the interplay of the spin frustration and the quantum fluctuation produce the two-step phase transitions in CsFeBr$_3$. Microscopic measurements as neutron scattering and NMR are necessary to elucidate the nature of the phase transitions.

Finally we discuss the magnetization process for $H\parallel c$. Assuming that the spin structure in the ordered state is an umbrella structure, in which the $xy$-components of spins form the 120$^{\circ}$ structure in the basal plane, we can calculate the magnetization curve for $T=0$ by the mean-field approximation shown in the Appendix. The thin solid line in Fig. 9 is the magnetization curve calculated with $g_{\parallel}=2.65$ and the magnetic parameters obtained by Dorner {\it et al}.\cite{Dorner1} (see Table I). It is noted that the ground state at zero field is gapless for the magnetic parameters obtained by Visser and Harrison.\cite{Visser1,Harrison} The dashed line is the $M-H$ curve obtained by neglecting the highest $|-1\,\rangle$ state. In Fig. 9, we also plotted the experimental result obtained at $T=1.6$ K, from which the magnetization due to the Van Vleck paramagnetism was subtracted. 

The saturation field $H_{\rm s}$ and the gap field $H_{\rm g}$ are obtained to be $H_{\rm s}=33.7$ T and $H_{\rm g}=1.8$ T, which agree roughly with the experimental saturation field and the gap field. In more detail the calculated saturation field $H_{\rm s}$ is somewhat larger than the experimental value of $H_{\rm s}=28$ T, which is expected for $T\rightarrow 0$, while the calculated gap field $H_{\rm g}$ is somewhat smaller than the experimental value of $H_{\rm g}=2.89$ T. The exchange parameters should be reduced slightly against the single ion anisotropy $D$ to reproduce the experimental saturation field and the gap field within the framework of the mean-field approximation. 

The calculated $M-H$ curve is the convex function for $H_{\rm g}<H<H_{\rm s}$ and exhibits remarkable rounding in the low field region. This behavior is ascribed to the contribution of the $|-1\,\rangle$ state. On the other hand, the magnetization curve observed in CsFeBr$_3$ is almost linear for $H_{\rm c1}<H<21$ T, and its slope increases steeply for $H>21$ T (see also $dM/dH$ versus $H$ in Fig. 7(b)). This behavior is characteristic of the $S=1$ one-dimensional antiferromagnetic system.\cite{Sakai,Okunishi} The experimental result indicates that the magnetization process in CsFeBr$_3$ is dominated by the spin correlation in the linear chain along the $c$-axis.

\section{Conclusions}
We have presented the results of magnetization and specific heat measurements on a pseudo spin-1 triangular antiferromagnet CsFeBr$_3$ with a singlet ground state due to the large single ion anisotropy. The specific heat measurements revealed that the field-induced magnetic ordering for $H\parallel c$ occurs in two steps. The phase diagram was determined, as shown in Figs. 4 and 8. The successive phase transitions cannot be understood within the framework of the mean-field theory. We suggest that the phase transitions are attributed to the interplay of the spin frustration and the quantum fluctuation originated in the singlet ground state triangular antiferromagnet.

\section*{Acknowledgements}
The authors would like to acknowledge K. Katsumata for allowing us access to the Mag Lab$^{\rm HC}$ calorimeter system in RIKEN. They also express their sincere thanks to M. Chiba, H. Shiba and K. Kindo for fruitful discussions and comments on the magnetic phase transitions in hexagonal AFeX$_3$ antiferromagnets and magnetization process in quasi-one-dimensional antiferromagnetic systems.

\section*{Appendix: Magnetization Curve for $H\parallel c$ at\\ \hspace{7.5cm}$T=0$}
Using the mean-field approximation, we calculate the magnetization curve at $T=0$ on the basis of the six-sublattice model. We assume that an umbrella spin structure is realized in the ordered state, for which all the sublattice spins are equivalent. We take the sublattices $1\sim 3$ in a $c$-plane and $4\sim 6$ in the next $c$-plane. We describe the basis state for the $n$-th sublattice spin as
$$\psi_n=|\,0\,\rangle \cos{\theta}+\left(\,|\,1\,\rangle \cos{\phi}+|-1\,\rangle \sin{\phi}\,\right)\sin{\theta}\,{\rm e}^{\frac{2\pi}{3}i(n-1)}\ , \eqno({\rm A}.1)$$
for $n=1\sim 3$, and
$$\psi_n=|\,0\,\rangle \cos{\theta}-\left(\,|\,1\,\rangle \cos{\phi}+|-1\,\rangle \sin{\phi}\,\right)\sin{\theta}\,{\rm e}^{\frac{2\pi}{3}i(n-1)}\ , \eqno({\rm A}.2)$$
for $n=4\sim 6$, where $|\,0\,\rangle$ and $|\pm1\,\rangle$ denote the spin states for $S^z=0$ and $\pm1$, respectively. Angles $\theta$ and $\phi$ were introduced to satisfy the normalization condition. The average values of spin operators are given by
$$\langle S_n^z\rangle =\sin^2{\theta}\cos{2\phi} , \qquad \langle \left(S_n^z\right)^2\rangle=\sin^2{\theta}\ , \eqno({\rm A}.3)$$
\begin{eqnarray}
\langle S_1^+\rangle &=& \langle S_1^-\rangle^*=-\langle S_4^+\rangle=-\langle S_4^-\rangle^* \nonumber\\
    &=& \sqrt{2}\cos{\theta}\sin{\theta}\,(\cos{\phi}+\sin{\phi})\ , \hspace{2.1cm}({\rm A}.4) \nonumber \\
\langle S_2^+\rangle &=& \langle S_2^-\rangle^*=-\langle S_5^+\rangle=-\langle S_5^-\rangle^* \nonumber \\
    &=& \sqrt{2}\cos{\theta}\sin{\theta}\,({\rm e}^{-\frac{2\pi}{3}i}\cos{\phi}+{\rm e}^{\frac{2\pi}{3}i}\sin{\phi})\ , \hspace{0.5cm}({\rm A}.5) \nonumber \\
\langle S_3^+\rangle &=& \langle S_3^-\rangle^*=-\langle S_6^+\rangle=-\langle S_6^-\rangle^* \nonumber \\
    &=& \sqrt{2}\cos{\theta}\sin{\theta}\,({\rm e}^{\frac{2\pi}{3}i}\cos{\phi}+{\rm e}^{-\frac{2\pi}{3}i}\sin{\phi})\ . \hspace{0.6cm}({\rm A}.6) \nonumber
\end{eqnarray}
Therefore, the energy per site $E$ is expressed by 
\begin{eqnarray*} 
E & = & D\sin^2{\theta}-g_{\parallel}\mu_{\rm B}H\sin^2{\theta}\cos{2\phi} \nonumber \\
  & & {} -2\left(2J_0^{\perp}+3J_1^{\perp}\right)\cos^2{\theta}\sin^2{\theta}(1+\sin{2\phi}) \nonumber \\
  & & {} +2\left(J_0^{\parallel}+3J_1^{\parallel}\right)\sin^4{\theta}\cos^2{2\phi}\ . \hspace{2.2cm}({\rm A}.7)
\end{eqnarray*}
Angles $\theta$ and $\phi$ can be determined by $\partial E/\partial \theta=0$ and $\partial E/\partial \phi=0$, which lead to
\begin{eqnarray*}
\sin^2{\theta}= \hspace{7cm}\\
\frac{g_{\parallel}\mu_{\rm B}H\cos{2\phi}-D+2\left(2J_0^{\perp}+3J_1^{\perp}\right)(1+\sin{2\phi})}{4\left[\left(2J_0^{\perp}+3J_1^{\perp}\right)(1+\sin{2\phi})+\left(J_0^{\parallel}+3J_1^{\parallel}\right)\cos^2{2\phi}\right]}\ , \\
({\rm A}.8) \\
\end{eqnarray*}
and
\begin{eqnarray*}
g_{\parallel}\mu_{\rm B}H\sin{2\phi}-2\left(2J_0^{\perp}+3J_1^{\perp}\right)\cos^2{\theta}\cos{2\phi} \hspace{2cm}\\
-2\left(J_0^{\parallel}+3J_1^{\parallel}\right)\sin^2{\theta}\sin{4\phi}=0\ , \hspace{0.5cm} ({\rm A}.9)
\end{eqnarray*}
for the ordered state. The magnetization curve is obtained by solving eqs. ({\rm A}.8) and ({\rm A}.9) self-consistently. In the saturated state, $\sin{\theta}=1$ and $\sin{\phi}=0$. Thus, the saturation field $H_{\rm s}$ is given by
$$g_{\parallel}\mu_{\rm B}H_{\rm s}=D+2\left(2J_0^{\perp}+3J_1^{\perp}+2J_0^{\parallel}+6J_1^{\parallel}\right)\ . \eqno({\rm A}.10)$$
At the gap field $H_{\rm g}$, $\sin{\theta}=0$. Substituting this condition into eqs. ({\rm A}.8) and ({\rm A}.9), we obtain 
$$g_{\parallel}\mu_{\rm B}H_{\rm g}=\sqrt{D^2-4D\left(2J_0^{\perp}+3J_1^{\perp}\right)}\ . \eqno({\rm A}.11)$$
If we neglect the highest $|-1\,\rangle$ state, {\it i.e.}, $\sin{\phi}\equiv 0$, the we obtain
$$g_{\parallel}\mu_{\rm B}H_{\rm g}=D-2\left(2J_0^{\perp}+3J_1^{\perp}\right)\ . \eqno({\rm A}.12)$$
From eq. ({\rm A}.11), the critical value of $D$, which separates the ground state natures in the absence of the magnetic field, is given by 
$$D_{\rm c}=4\left(2J_0^{\perp}+3J_1^{\perp}\right)\ . \eqno({\rm A}.13)$$
The ground state can have a magnetic ordering for $D<D_{\rm c}$ while it is a spin singlet for $D>D_{\rm c}$.

\end{document}